\documentclass[11pt]{article}
\usepackage[margin=1.1in]{geometry}
\usepackage[numbers,sort&compress]{natbib}

\usepackage{graphicx}
\usepackage{float}
\usepackage{hyperref}
\usepackage{epstopdf}
\usepackage{amsmath}
\usepackage{verbatim}
\usepackage[english]{babel}
\usepackage{multirow}
\renewcommand{\arraystretch}{1.3} % to make tables more readable
\usepackage{array}
\usepackage{tabularx}
\usepackage{longtable}
\usepackage{tikz}
\usetikzlibrary{shapes.geometric, arrows, positioning}

\newcolumntype{P}[1]{>{\raggedright\arraybackslash}p{#1}}
\newcolumntype{Q}{>{\raggedright\arraybackslash}X}

\newcommand{\MateraTableSetup}{\small\setlength{\tabcolsep}{6pt}\renewcommand{\arraystretch}{1.2}}
\newcommand{\MateraTH}[1]{\textbf{#1}}

\usepackage[utf8]{inputenc}
\usepackage{microtype}
\usepackage{listings}
\usepackage{xcolor}
 
\definecolor{codegreen}{rgb}{0,0.45,0.15}
\definecolor{codegray}{rgb}{0.75,0.75,0.75}
\definecolor{codepurple}{rgb}{0.2235,0.1608,0.4941}
\definecolor{codeblue}{rgb}{0.0,0.2,0.6}
\definecolor{codebrown}{rgb}{0.6,0.2,0.0}
\definecolor{codeblack}{rgb}{0.15,0.15,0.15}
\definecolor{backcolour}{rgb}{0.98,0.98,0.98}

\lstdefinestyle{mystyle}{
    language=Python,
    backgroundcolor=\color{backcolour},
    basicstyle=\scriptsize\ttfamily\color{codeblack},
    commentstyle=\color{codegreen}\itshape,
    keywordstyle=\color{magenta}\bfseries,
    numberstyle=\tiny\color{codegray},
    stringstyle=\color{codebrown},
    identifierstyle=\color{codeblack},
    emph={Q3ASECalculator,run_relax_workflow,run_band_structure_workflow,run_vc_relax_workflow,run_scf,BFGS,bulk,read},
    emphstyle=\color{codepurple}\bfseries,
    breaklines=true,
    captionpos=b,
    keepspaces=true,
    numbers=left,
    numbersep=4pt,
    showstringspaces=false,
    xleftmargin=0.1\linewidth,
    xrightmargin=0.1\linewidth,
    frame=single,
    framerule=0pt,
    rulecolor=\color{backcolour},
    framesep=2pt,
    aboveskip=10pt,
    belowskip=10pt,
    columns=fullflexible,
    tabsize=2
}
\makeatletter
\renewcommand{\@fnsymbol}[1]{\@arabic{#1}}
\makeatother

\usepackage{titling}
\pretitle{\begin{center}\LARGE}
\posttitle{\par\end{center}\vspace{0.6em}}
\preauthor{\begin{center}\large\lineskip 0.4em\begin{tabular}[t]{c}}
\postauthor{\end{tabular}\par\end{center}}
\predate{}
\postdate{}

\title{Quantum 3xplorer: AI-Native Engine for \\ Next-Generation Quantum Materials Modeling}

\author{
    Vsevolod Biryukov\textsuperscript{1}
    \and Sergey V. Barabash\textsuperscript{1}
    \and Pranab Kumar Das\textsuperscript{1}
    \and Jack Deslippe\textsuperscript{2}
    \and Mauro Del Ben\textsuperscript{3}
    \and Felipe H. da Jornada\textsuperscript{4}
    \and Yuming Shi\textsuperscript{4}
    \and Johannes Voss\textsuperscript{5}
    \and Zhenglu Li\textsuperscript{6}
    \and Kamal Choudhary\textsuperscript{7}
    \and Ruth Pachter\textsuperscript{8}
    \and Steven G. Louie\textsuperscript{9,10}
    \and Marvin L. Cohen\textsuperscript{9}
    \and Timur Bazhirov\textsuperscript{1,7,*}
}
\date{}

\begin{document}
    \maketitle
    \vspace{-1em}
    \footnotetext[1]{Exabyte Inc. (Mat3ra.com), Walnut Creek, CA 94596, USA}
    \footnotetext[2]{National Energy Research Scientific Computing Center (NERSC), Lawrence Berkeley National Laboratory, Berkeley, CA 94720, USA}
    \footnotetext[3]{Applied Mathematics and Computational Research Division, Lawrence Berkeley National Laboratory, Berkeley, CA 94720, USA}
    \footnotetext[4]{Department of Materials Science and Engineering, Stanford University, Stanford, CA 94305, USA}
    \footnotetext[5]{SUNCAT Center for Interface Science and Catalysis, SLAC National Accelerator Laboratory, Menlo Park, CA 94025, USA}
    \footnotetext[6]{Mork Family Department of Chemical Engineering and Materials Science, University of Southern California, Los Angeles, CA 90089, USA}
    % \footnotetext[7]{NVIDIA Corporation, Santa Clara, CA 95051, USA}
    \footnotetext[7]{Department of Materials Science and Engineering, Johns Hopkins University, Baltimore, MD 21218, USA}
    \footnotetext[8]{Air Force Research Laboratory, Wright-Patterson Air Force Base, OH 45433, USA}
    \footnotetext[9]{Department of Physics, University of California at Berkeley, Berkeley, CA 94720, USA}
    \footnotetext[10]{Materials Sciences Division, Lawrence Berkeley National Laboratory, Berkeley, CA 94720, USA}
    {\renewcommand{\thefootnote}{*}\footnotetext{Corresponding author: timur@mat3ra.com}}
    \begin{abstract}
    Quantum-mechanical approaches, such as density functional theory (DFT), underpin much of modern computational materials science and are the primary source of data for AI-driven discovery. Yet the established electronic-structure codes today represent decades-old software architectures, making new physical models slow and costly to implement, validate, and deploy on emerging hardware. We present Quantum 3xplorer (Q3, pronounced ``Quantum Explorer''), an AI-native quantum simulation engine written from scratch by AI coding agents under the direction and supervision of domain scientists. As a first and demanding validation case, we implement plane-wave pseudopotential DFT end to end: norm-conserving, ultrasoft, and projector-augmented-wave formalisms, spin polarization, DFT+U, and dispersion corrections, together with forces, stresses, structural relaxation, and electronic-structure analysis. Q3 reproduces established reference implementations, runs faster on a single core in representative benchmarks, and scales across GPU-accelerated supercomputers. It was developed in weeks rather than years. Its layered, physics-first architecture separates abstract quantum-mechanical concepts from their numerical representations, making its framework extendable beyond plane waves to other basis sets and levels of theory. Its inputs and outputs follow structured, schema-validated data standards, making every result AI-ready. Q3 is open source under Apache 2.0 and built to be reused, extended, and improved by the research community. Q3 points to a paradigm shift in how computational quantum science is done. Scientists define the physics, formulate the models, and set the criteria for validation; AI agents write, test, and optimize the code. The decades-long path from a new theory to a trusted, high-performance implementation can thus collapse to days, significantly accelerating scientific discovery.
\end{abstract}

    \section{Introduction}
\label{sec:introduction}

    First-principles materials simulation based on Density Functional Theory (DFT) \cite{hohenberg-kohn1964DFT, kohn1965self} underpins progress in areas central to national competitiveness, including energy storage, semiconductor electronics, and quantum materials. In recent years, these simulations have also become the primary \textit{data engine} for the rapid growth of artificial intelligence in the physical sciences. High-quality datasets of ground-truth energies, forces, and electronic properties are essential for training machine-learned interatomic potentials and driving large-scale, AI-accelerated materials discovery \cite{merchant2023gnome, jain2013materialsproject, saal2013openQMD, curtarolo2012aflowlib, nomad, pizzi2016aiida}.

    The workhorse engines of the community, such as Quantum ESPRESSO \cite{QE2009mainReference, giannozzi2017qe} and VASP \cite{kresse1996efficient}, represent decades of cumulative theoretical, algorithmic, and numerical innovation in Fortran. These foundational packages remain the gold-standard reference implementations of solid-state physics, providing both the theoretical corpus and the rigorous validation benchmarks upon which any new simulation software must build. Because their architectures predated modern machine learning frameworks, heterogeneous GPU accelerators, and AI coding assistants, coupling monolithic procedural codebases directly to end-to-end differentiable pipelines or automated agentic development workflows requires substantial adaptation. Recent efforts have begun addressing aspects of this challenge through domain-specific C++ GPU codes such as SIRIUS \cite{zhang2023sirius} and high-level Julia frameworks such as DFTK \cite{herbst2021dftk}. Complementing these advances, we explore a new software paradigm centered on an \textit{AI-native} lifecycle: a simulation engine designed to be understandable for AI, written by AI, maintained by AI, and reviewed and validated by humans.

    We present \textbf{Quantum 3xplorer (Q3)}, a modern, from-scratch, AI-native quantum simulation engine designed to address these challenges. Q3 is not intended to replace mature, feature-rich codes, but rather to provide a clean, extensible, and high-performance core designed for the workflows of the next decade. The novelty of Q3 lies not in the underlying equations of DFT, which are well established, but in \textit{how} the engine is conceived, built, and sustained:

    \begin{itemize}
        \item \textbf{AI-Native by Design}: Q3 is built around a human-AI co-development paradigm: understandable for AI, written and maintained by AI coding agents, and reviewed and validated by human scientists. Its strictly layered C++17 architecture, one-class-per-file modularity, and avoidance of domain-specific abbreviations serve this concept by making every component self-documenting and legible within an AI agent's context window.
        \item \textbf{Structured and Interoperable}: Built on the open-source CAMSAI data standards \cite{camsai-standards, mat3ra-esse}, Q3 utilizes schema-validated JSON and YAML for inputs and outputs alongside drop-in parsers for community formats, ensuring that calculation inputs and results are machine-readable and reproducible by construction.
        \item \textbf{MPI and GPU Parallel Acceleration}: Q3 features production-grade MPI k-point pool and band-group parallel scaling and native GPU acceleration across four vendor backends (NVIDIA CUDA and Apple Metal, validated; AMD HIP and Intel SYCL, compiling in CI), enabling high-efficiency scaling across distributed nodes and GPU accelerators on leadership systems like Perlmutter at NERSC \cite{nersc2021perlmutter}.
        \item \textbf{Method-Agnostic Core}: The engine's architecture separates abstract quantum-mechanical operators and physical models from concrete numerical representations and solvers \cite{EB-ARX-2021}. While the current implementation focuses on plane-wave pseudopotential DFT, the core is designed to support alternative discretizations and many-body methods without rewriting the upper property and workflow layers.
        \item \textbf{Differentiable Simulation}: Q3 is architected to support a differentiable interface for machine learning applications, producing exact analytical gradients for backpropagation in neural network potential training loops.
    \end{itemize}

    In its current state, Q3 implements the core plane-wave pseudopotential pipeline \cite{mlcohen1979pseudopotentialDFT}, including norm-conserving (NC) \cite{troullier1991nc, oncvpsp2013}, ultrasoft (US) \cite{vanderbilt1990uspp}, and projector augmented wave (PAW) \cite{blochl1994paw} pseudopotentials, self-consistent field (SCF) iteration, forces, stress, and electronic structure properties. Beyond the core DFT pipeline, Q3 supports collinear spin-polarized LSDA calculations \cite{vonbarth1972lsda}, DFT-D3(BJ) van der Waals dispersion corrections \cite{grimme2010dftd3, grimme2011dftd3bj}, Hubbard DFT+U in the Dudarev formulation \cite{dudarev1998dftu}, charged-cell electrostatic corrections \cite{makov1995makovpayne, martyna1999martinatuckerman}, projected density of states with L\"owdin population analysis \cite{lowdin1950population}, and variable-cell structural relaxation via ASE cell filters \cite{larsen2017atomic}. Q3 provides drop-in compatibility with Quantum ESPRESSO input files and UPF pseudopotentials, alongside an engine-neutral build-time add-on registry architecture for integrating additional external input, dataset, and output formats, enabling seamless migration of existing workflows. In this paper, we describe the design principles of Q3, its architectural foundations, and its validation against established reference implementations. We demonstrate that Q3 achieves competitive performance while maintaining high numerical accuracy, proving that an AI-native approach to scientific software development is both feasible and advantageous.

    The remainder of this manuscript is organized as follows. Section~\ref{sec:methodology} outlines the layered architecture, MPI scaling framework, and GPU execution backends. Section~\ref{sec:results} summarizes validation results, including accuracy comparisons, single-core speedups, MPI pool parallel scaling, and GPU acceleration benchmarks. Section~\ref{sec:discussion} discusses the implications of AI-native design for materials discovery, code provenance and open-science governance, and future directions. Section~\ref{sec:conclusion} concludes the work.

    \section{Methodology}
\label{sec:methodology}

    \subsection{System Architecture and Design Principles}
    \label{subsec:methodology-architecture}

        Quantum 3xplorer (Q3) is built on a strictly layered C++17 architecture designed to decouple physical concepts from computational implementation. The codebase is organized into six functional layers (0--5), where each layer only depends on the layers below it. This architecture enforces strict modularity, preventing cyclic dependencies and ensuring that the engine is highly extensible.

        Figure~\ref{fig:q3-architecture} illustrates the layout of the layers, the vertical File I/O slice, and optional build-time add-ons:

        \begin{figure}[H]
            \centering
            \resizebox{\linewidth}{!}{%
            \begin{tikzpicture}[
                layer/.style={rectangle, draw=blue!80!black, fill=blue!5, thick, minimum width=7.5cm, minimum height=1.0cm, align=center, rounded corners},
                io/.style={rectangle, draw=green!60!black, fill=green!5, thick, minimum width=2.6cm, minimum height=6.0cm, align=center, rounded corners},
                addon/.style={rectangle, draw=orange!80!black, fill=orange!5, thick, dashed, minimum width=2.9cm, minimum height=6.0cm, align=center, rounded corners},
                arrow/.style={-latex, thick, draw=gray!80}
            ]
                % Layers
                \node[layer] (l5) {Layer 5: Properties \\ \footnotesize (Forces, Stress Tensor, Results Serialization)};
                \node[layer, below=0.3cm of l5] (l4) {Layer 4: Computation \\ \footnotesize (SCF Iterator, Mixers, Eigensolvers, MPI, GPU)};
                \node[layer, below=0.3cm of l4] (l3) {Layer 3: Representation \\ \footnotesize (Plane-wave Basis, G-vectors, FFT, Pseudopotentials)};
                \node[layer, below=0.3cm of l3] (l2) {Layer 2: Quantum System \\ \footnotesize (Abstract wavefunctions, densities, potentials)};
                \node[layer, below=0.3cm of l2] (l1) {Layer 1: Structure \\ \footnotesize (Lattice, Atomic positions, Space group symmetry)};
                \node[layer, below=0.3cm of l1] (l0) {Layer 0: Core \\ \footnotesize (Types, Math utilities, Constants, Logger, GPU Device Abstraction)};
                
                % Vertical I/O Slice
                \node[io, right=0.55cm of l3] (io) {File I/O Slice \\ \footnotesize (YAML/QE Input, \\ \footnotesize UPF Readers, \\ \footnotesize Checkpoints, \\ \footnotesize JSON Output)};

                % Optional Build-Time Add-Ons
                \node[addon, right=0.45cm of io] (addon) {Build-Time Add-Ons \\ \footnotesize \texttt{src/addons/<name>/} \\ \footnotesize (Optional Format, \\ \footnotesize Dataset \& Convention \\ \footnotesize Plug-In Registries)};
                
                % Dependency Arrows
                \draw[arrow] (l5) -- (l4);
                \draw[arrow] (l4) -- (l3);
                \draw[arrow] (l3) -- (l2);
                \draw[arrow] (l2) -- (l1);
                \draw[arrow] (l1) -- (l0);
                
                % I/O Connections
                \draw[arrow, <->] (io.west) -- ++(-0.3,0) |- (l5.east);
                \draw[arrow, <->] (io.west) -- ++(-0.3,0) |- (l4.east);
                \draw[arrow, <->] (io.west) -- (l3.east);
                \draw[arrow, <->] (io.west) -- ++(-0.3,0) |- (l2.east);
                \draw[arrow, <->] (io.west) -- ++(-0.3,0) |- (l1.east);

                % Add-On One-Way Registration
                \draw[arrow] (addon.west) -- (io.east);
            \end{tikzpicture}%
            }
            \caption{Layered system architecture of Quantum 3xplorer. Each layer only depends on the layers beneath it, while the File I/O slice acts as a vertical integration pathway across all layers. Optional build-time add-ons (\texttt{src/addons/<name>/}) register external format, dataset, and energy-reference adapters into core registries via a strict one-way dependency boundary.}
            \label{fig:q3-architecture}
        \end{figure}
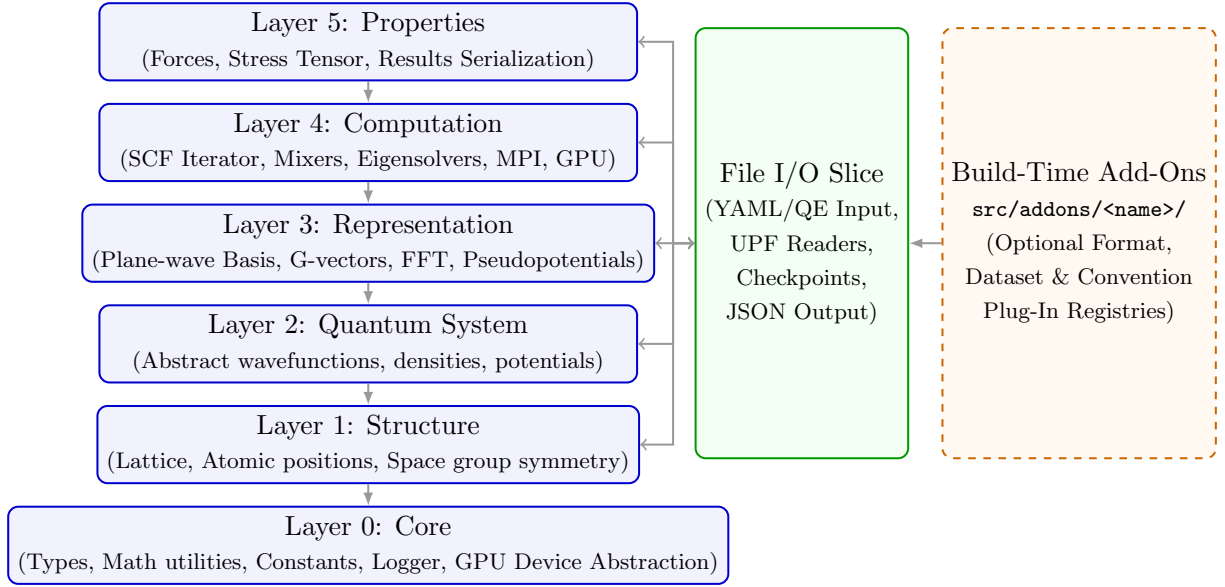

        The layers are defined as follows:
        \begin{itemize}
            \item \textbf{Layer 0 (Core)}: Provides fundamental types (Real, Complex, Vector3D, Matrix3x3), math utilities (radial integration, spherical harmonics), GPU device abstractions (\texttt{GPUBLAS}, \texttt{GPUEigensolver}, \texttt{GPUDeviceBackend}), and logging infrastructure.
            \item \textbf{Layer 1 (Structure)}: Handles the crystal structure, including lattice vectors, atomic positions, and space-group symmetry detection via SPGlib \cite{spglib2018}. It also manages k-point grid generation using the Monkhorst-Pack scheme \cite{monkhorst1976kpoints}.
            \item \textbf{Layer 2 (Quantum System)}: Defines abstract quantum-mechanical interfaces for wavefunctions, electron densities, and potentials, independent of the basis set discretization.
            \item \textbf{Layer 3 (Representation)}: Implements the concrete plane-wave representation, including basis sets, Fast Fourier Transform (FFT) grids (\texttt{FFTGrid}, \texttt{GPUFFTGrid}), G-vector management, and pseudopotential handling. Norm-Conserving (NC) \cite{troullier1991nc, oncvpsp2013}, Ultrasoft (US) \cite{vanderbilt1990uspp}, and Projector Augmented Wave (PAW) \cite{blochl1994paw} formalisms are supported, with built-in readers for UPF v1/v2 formats and extensible registries for additional dataset and one-center energy-reference conventions.
            \item \textbf{Layer 4 (Computation)}: Contains the core algorithms for solving the Kohn-Sham equations, including SCF iterators, eigensolvers (Davidson \cite{davidson1975} and Direct/LAPACK \cite{lapack1999}), density mixers (Broyden \cite{broyden1965} and Anderson \cite{anderson1965}), MPI k-point pool parallel orchestrators, and GPU execution backends. This layer also hosts the DFT-D3(BJ) dispersion calculator \cite{grimme2010dftd3, grimme2011dftd3bj} and the Dudarev DFT+U implementation \cite{dudarev1998dftu}.
            \item \textbf{Layer 5 (Properties)}: Implements post-SCF calculations for Hellmann-Feynman forces \cite{feynman1939forces, pulay1969forces} and the Nielsen-Martin stress tensor \cite{nielsen1985stress}, projected density of states with L\"owdin population analysis \cite{lowdin1950population}, and manages the serialization of results to structured formats. Electrostatic corrections for charged cells (Makov-Payne \cite{makov1995makovpayne} and Martyna-Tuckerman \cite{martyna1999martinatuckerman}) are also evaluated at this layer.
        \end{itemize}

        This layered decomposition aligns directly with the open-source CAMSAI schema hierarchy \cite{camsai-standards, mat3ra-esse, EB-ARX-2021}, mapping Layer~1 to the structural \textit{Material} definition, Layers~2 and~4 to the physical \textit{Model} and numerical \textit{Method} specifications, and Layer~5 to the output \textit{Property} schemas.

        A ``vertical slice'' for I/O handles built-in file operations (CAMSAI standards YAML and Quantum ESPRESSO namelist inputs, structured JSON outputs, UPF v1/v2 pseudopotential readers, Gaussian cube and XSF volumetric exports, and binary checkpoints) by interacting with all relevant layers. To accommodate external ecosystems, including proprietary simulation packages, commercially licensed pseudopotential libraries, or institution-specific file formats whose distribution or licensing terms differ from the Apache-2.0 core, Q3 implements a modular build-time add-on architecture (\texttt{src/addons/<name>/}, with mirrored \texttt{tests/} and \texttt{examples/} subtrees). Add-ons act as self-contained vertical slices that can be maintained in independent (including private) repositories and overlaid onto a matching core checkout at configure time. Each add-on reaches the core exclusively through five engine-neutral registries: input deck parsers, pseudopotential dataset readers, PAW one-center energy-reference conventions (backed by polymorphic dataset extension slots on \texttt{PawPseudopotential}), post-SCF companion output writers, and volumetric field formats. Dependency flow is strictly one-way: add-ons may include core headers, but no core file ever includes, links to, branches on, or names an add-on. When an add-on directory is present, CMake automatically compiles it, registers its unit and validation test suites, and records its identity and version in \texttt{q3 --version} and the output \texttt{provenance.generator.addons} metadata; when absent, the core builds and passes all verification gates as a standalone open-source package.

        Q3 provides eight distinct executable entry points: \texttt{q3} (serial), \texttt{q3\_mpi} (MPI pool/band parallelism), \texttt{q3\_gpu} (Apple Metal), \texttt{q3\_gpu\_cuda} (NVIDIA CUDA), \texttt{q3\_mpi\_gpu\_cuda} (NVIDIA CUDA across several GPUs, one device and one k-point pool per MPI rank), \texttt{q3\_gpu\_hip} (AMD ROCm), \texttt{q3\_gpu\_sycl} (Intel oneAPI), and \texttt{q3-util} (post-processing utility). The \texttt{q3-util} utility provides subcommands for converting converged mean-field records to BerkeleyGW \cite{deslippe2012berkeleygw} formats (\texttt{WFN}, \texttt{RHO}, \texttt{vxc.dat}) for post-DFT GW/BSE calculations \cite{hybertsen1986gw, rohlfing2000bse}, and for evaluating Wannier-interpolated \cite{marzari2012wannier} band structures and densities of states at arbitrary k-points from a saved interpolation record.

        While all Kohn-Sham operators, pseudopotential projectors (NC, US, and PAW), Davidson eigensolvers, density mixers, Hellmann-Feynman force and Nielsen-Martin stress calculators, DFT+U, DFT-D3(BJ), and Wannier interpolation routines are implemented from scratch in Q3, the engine builds upon established open-source community libraries for standard mathematical kernels, symmetry analysis, exchange-correlation primitives, and structured serialization. Table~\ref{tab:open-source-dependencies} summarizes these open-source dependencies and their roles across the layered architecture.

        \begin{table}[H]
            \centering
            \MateraTableSetup
            \caption{Open-source libraries and community standards leveraged across the Q3 architecture alongside first-party electronic-structure implementations.}
            \label{tab:open-source-dependencies}
            \begin{tabularx}{\linewidth}{|l|l|X|}
                \hline
                \MateraTH{Component / Library} & \MateraTH{Layer} & \MateraTH{Role in Q3} \\
                \hline
                BLAS / LAPACK \cite{lapack1999} & Layer 0, 4 & Dense linear algebra primitives (\texttt{dgemm}/\texttt{zgemm}, Cholesky QR, subspace eigensolvers) \\ \hline
                FFTW3 \cite{fftw2005} / PocketFFT / cuFFT / vDSP & Layer 3 & 3D Fast Fourier Transforms between reciprocal ($\mathbf{G}$) and real ($\mathbf{r}$) space \\ \hline
                Vendor GPU SDKs (CUDA, Metal, HIP, SYCL) & Layer 0, 3, 4 & Device memory management, batched FFTs, and accelerated BLAS/eigensolver backends \\ \hline
                SPGlib \cite{spglib2018} & Layer 1 & Crystal space-group symmetry detection and irreducible Brillouin-zone k-point reduction \\ \hline
                LibXC \cite{libxc2018} & Layer 3, 4 & Pointwise evaluation of LDA and GGA exchange-correlation energy densities and potentials \\ \hline
                CAMSAI standards \cite{camsai-standards, mat3ra-esse} / \texttt{nlohmann\_json} / \texttt{yaml-cpp} & Layer 5, I/O & Schema-validated structured input/output serialization and data interoperability \\ \hline
                ASE \cite{larsen2017atomic} / \texttt{pybind11} & Python API & In-memory C++ bindings (\texttt{q3\_core}), geometry optimization filters, and workflow orchestration \\ \hline
                \texttt{spdlog} / GoogleTest & Layer 0, Tests & MPI-aware structured logging and automated C++ unit/integration test suite (1,637 tests) \\ \hline
            \end{tabularx}
        \end{table}

    \subsection{AI-Native Design and Conventions}
    \label{subsec:methodology-ai-native}

        Q3 is designed from the ground up around an AI-native software engineering paradigm: \textit{understandable for AI, written by AI, maintained by AI, and reviewed and validated by humans}. This concept is realized through four foundational conventions:

        \begin{itemize}
            \item \textbf{Explicit Naming}: The codebase avoids domain-specific abbreviations. Variable and class names are fully spelled out (e.g., \texttt{number\_of\_plane\_waves} instead of \texttt{npw}) to ensure the code is self-documenting for both human developers and LLMs.
            \item \textbf{One Class Per File}: Each header file contains exactly one class definition, mirroring the project structure and making it easy for AI agents to locate and modify specific components.
            \item \textbf{Strategy and Factory Patterns}: Interchangeable components such as exchange-correlation functionals, mixers, and diagonalizers are implemented via clean interfaces, allowing AI agents to extend the engine with new features without breaking existing logic.
            \item \textbf{Structured Data Standards}: Q3 uses the CAMSAI standards \cite{camsai-standards, mat3ra-esse} for all inputs and outputs. This ensures that the engine is naturally integrated into automated, data-driven pipelines.
        \end{itemize}

    \subsection{Parallel Execution: MPI and GPU Infrastructure}
    \label{subsec:methodology-implementation}

        Q3 implements a multi-level parallel computational architecture designed for HPC scalability and performance portability:
        
        \begin{itemize}
            \item \textbf{MPI K-Point Pool and Band-Group Parallelism}: For multi-k-point calculations, Q3 partitions the irreducible Brillouin zone across MPI communicators (pools). Each MPI rank owns a subset of k-points and independently evaluates wavefunctions and Hamiltonian projections. Within each pool, ranks divide reciprocal space by FFT stick (keeping the $\Gamma$-only half sphere), so that per-rank memory falls with the rank count; an alternative band-distribution mode (\texttt{--distribute-bands}) instead splits the Kohn-Sham states across the pool, which returns more speed for $\Gamma$-only and few-k-point systems. Collective communication (\texttt{MPI\_Allreduce}) is used exclusively for global charge density accumulation, Fermi energy bisection, and total energy component reductions, guaranteeing exact topology-invariant energy matching across rank counts.
            \item \textbf{GPU Hardware Abstraction}: Layer 0 defines vendor-agnostic GPU interfaces (\texttt{GPUBLAS}, \texttt{GPUFFTGrid}, \texttt{GPUEigensolver}, and \texttt{GPUDeviceBackend}). Vendor backends are conditionally compiled for NVIDIA CUDA (cuBLAS, cuFFT, cuSOLVER) and Apple Metal (Accelerate / vDSP), with AMD ROCm (HIP) and Intel SYCL backends compiling in CI.
            \item \textbf{Persistent GPU Buffers \& Batched cuFFT}: To minimize host-to-device allocation latencies, device memory is pre-allocated in RAII wrappers (\texttt{GPUDeviceBuffer}). In Layer 3, Hamiltonian application is batched across all bands using \texttt{cufftPlanMany}, replacing single-band transforms with a single batched kernel launch.
            \item \textbf{Mixed-Precision Policy}: Through \texttt{PlanewavePrecisionPolicy}, Q3 supports full double-precision (FP64) and mixed-precision (FP32 wavefunction FFTs with FP64 energy accumulation) execution, providing acceleration while maintaining sub-$\mu$Ry accuracy.
            \item \textbf{Exchange-Correlation \& Smearing}: Integration with LibXC \cite{libxc2018} provides access to LDA and GGA functionals. Metallic systems are supported using Gaussian \cite{fu1983gaussian}, Methfessel-Paxton \cite{methfessel1989smearing}, Marzari-Vanderbilt cold smearing \cite{marzari1999smearing}, or Fermi-Dirac occupations \cite{mermin1965}.
        \end{itemize}

        The engine is continuously validated against established plane-wave DFT codes. A comprehensive suite of over 1,800 automated tests (including 1,637 GTest unit and integration tests, MPI scaling checks, CUDA CI workflows, 218 Python validation tests, and 32 regression tests) ensures numerical correctness across semiconductors, metals, and magnetic systems with NC, US, and PAW pseudopotentials. Q3 also supports collinear spin-polarized LSDA calculations with per-species and per-atom starting magnetizations, enabling ferromagnetic and antiferromagnetic ground-state studies.

    \subsection{Python Interface and Workflow Automation}
    \label{subsec:methodology-python}

        To support automated materials modeling, high-throughput workflows, and downstream integration with machine learning potential training, Q3 provides a Python layer built on modern interoperability standards. The Python package exposes both an ASE-compliant calculator interface (\texttt{Q3ASECalculator}) \cite{larsen2017atomic} and high-level workflow orchestrators (\texttt{q3.workflows}).

        Through the ASE calculator, Q3 evaluates total energies, Hellmann-Feynman atomic forces, and the Nielsen-Martin stress tensor directly on ASE \texttt{Atoms} objects, enabling seamless coupling with standard optimization engines (such as BFGS, LBFGS, and Frechet/ExpCellFilter for variable-cell relaxation). Multi-step simulation pipelines, such as self-consistent field (SCF) calculations followed by non-self-consistent field (NSCF) band-structure or density-of-states evaluations, are automated through modular workflow functions that manage parameter propagation and structured data extraction. Listing~\ref{lst:q3-workflow} provides a representative Python workflow demonstrating both paradigms: relaxing a crystal geometry via the ASE calculator and evaluating band structures along a high-symmetry path.

        \noindent\begin{minipage}{\linewidth}
\begin{lstlisting}[language=Python, firstline=7, caption={Representative Python workflow in Q3: ionic relaxation via the ASE calculator interface and automated band structure calculation.}, label={lst:q3-workflow}]
from ase.build import bulk
from ase.optimize import BFGS

from q3.calculators import Q3ASECalculator
from q3.workflows import run_band_structure_workflow

# 1. Structure definition & calculator attachment
atoms = bulk("Si", "diamond", a=5.43)
atoms.calc = Q3ASECalculator(
    scf_input_path="examples/total_energy--semiconductor--silicon/nc.yml"
)

# 2. Evaluate total energy and Hellmann-Feynman forces
potential_energy = atoms.get_potential_energy()
forces = atoms.get_forces()

# 3. Geometry relaxation with ASE BFGS optimizer
optimizer = BFGS(atoms, trajectory="silicon_relaxation.traj")
optimizer.run(fmax=0.01)

# 4. Multi-step band structure calculation (SCF + NSCF)
band_structure_results = run_band_structure_workflow(
    scf_input_path="examples/total_energy--semiconductor--silicon/nc.yml",
    band_path=["G", "X", "W", "K", "G", "L"],
    lattice_type="FCC",
    points_per_segment=30,
)
\end{lstlisting}
        \end{minipage}

    \section{Results}
\label{sec:results}

    \subsection{Validation and Accuracy}
    \label{subsec:results-accuracy}

        The numerical accuracy of Quantum 3xplorer (Q3) is continuously validated against an established plane-wave DFT reference implementation \cite{QE2009mainReference} and internal finite-difference consistency gates across three validation tiers: bulk crystalline materials (Section~\ref{subsubsec:results-accuracy-bulk}), magnetic surfaces with chemisorbed adsorbates and core-level excitations (Section~\ref{subsubsec:results-accuracy-surfaces}), and post-DFT many-body perturbation theory workflows (Section~\ref{subsubsec:results-accuracy-beyond-dft}). Across these benchmarks, calculations employ norm-conserving (NC) \cite{troullier1991nc}, ultrasoft (US) \cite{vanderbilt1990uspp}, and projector augmented wave (PAW) \cite{blochl1994paw} pseudopotentials with LDA and GGA functionals (PBE \cite{perdew1996generalized} and RPBE \cite{HammerPRB1999}) via LibXC \cite{libxc2018}.

        \subsubsection{Bulk Materials}
        \label{subsubsec:results-accuracy-bulk}

        The primary bulk benchmark configurations are defined as follows:
        \begin{itemize}
            \item \textbf{Silicon Primitive Cell}: Diamond cubic structure (2 atoms/cell), $8\times8\times8$ Monkhorst-Pack k-point grid, with wavefunction kinetic energy cutoffs of 48~Ry (NC) and 40~Ry (US). The corresponding charge density cutoffs are 192~Ry (NC) and 200~Ry (US). No smearing is applied.
            \item \textbf{Copper Primitive Cell}: FCC structure (1 atom/cell), $8\times8\times8$ Monkhorst-Pack k-point grid, and Gaussian smearing with a width of $\sigma = 0.02$~Ry. The kinetic energy cutoffs are set to 120~Ry (NC) for wavefunctions and 480~Ry (NC) for charge density, and 40~Ry (US) for wavefunctions and 200~Ry (US) for charge density.
            \item \textbf{Supercells}: For silicon, a $3\times3\times3$ supercell containing 54 atoms is used. For copper, a $3\times3\times3$ supercell containing 27 atoms is used. All supercell calculations are performed at the Gamma-point ($\Gamma$-only) with the SCF convergence threshold set to $10^{-8}$~Ry.
            \item \textbf{Magnetic Systems}: Iron BCC (ferromagnetic), nickel FCC (ferromagnetic), and chromium BCC (antiferromagnetic) are computed using collinear LSDA with per-species starting magnetizations.
        \end{itemize}

        Table~\ref{tab:accuracy-comparison} summarizes the agreement between Q3 and the reference implementation for total energies and band structure eigenvalues. In all non-magnetic cases, Q3 reproduces the reference total energies to within 0.014~mRy, with primitive cell energies often agreeing to 0.0002~mRy (0.1--0.2~$\mu$Ry per atom). Band energies show mean absolute differences at the sub-meV level. The spin-polarized LSDA results further validate Q3's magnetic implementation: Fe BCC-FM agrees to 0.11~mRy in total energy and 0.005~$\mu_B$ in magnetization, Ni FCC-FM to 0.013~mRy and 0.005~$\mu_B$, and Cr BCC-AFM to 0.25~mRy. DFT-D3(BJ) dispersion on graphite reproduces the interlayer equilibrium spacing to 0.0001~\AA{} and binding energy to 0.54~meV/atom versus the reference.

        \begin{table}[H]
            \centering
            \MateraTableSetup
            \caption{Q3 vs.\ Reference Implementation \cite{QE2009mainReference} accuracy (PBE, $8\times8\times8$ SCF grid for Si/Cu; spin-polarized LSDA for Fe/Ni/Cr).}
            \label{tab:accuracy-comparison}
            \begin{tabularx}{\linewidth}{|l|l|Q|Q|Q|}
                \hline
                \MateraTH{System} & \MateraTH{PP} & \MateraTH{SCF Energy $|\Delta E|$} & \MateraTH{Band mean $|\Delta E|$} & \MateraTH{Band max $|\Delta E|$} \\
                \hline
                Si & NC & 0.0002 mRy & 0.03 meV & 0.2 meV \\ \hline
                Si & US & 0.0002 mRy & 0.03 meV & 0.4 meV \\ \hline
                Cu & NC & 0.0002 mRy & 0.4 meV & 0.8 meV \\ \hline
                Cu & US & 0.014 mRy  & 0.05 meV & 0.2 meV \\ \hline
                Fe BCC-FM & NC & 0.11 mRy & --- & --- \\ \hline
                Ni FCC-FM & NC & 0.013 mRy & --- & --- \\ \hline
                Cr BCC-AFM & NC & 0.25 mRy & --- & --- \\ \hline
            \end{tabularx}
        \end{table}

\begin{figure}[H]
            \centering
            \includegraphics[width=\textwidth]{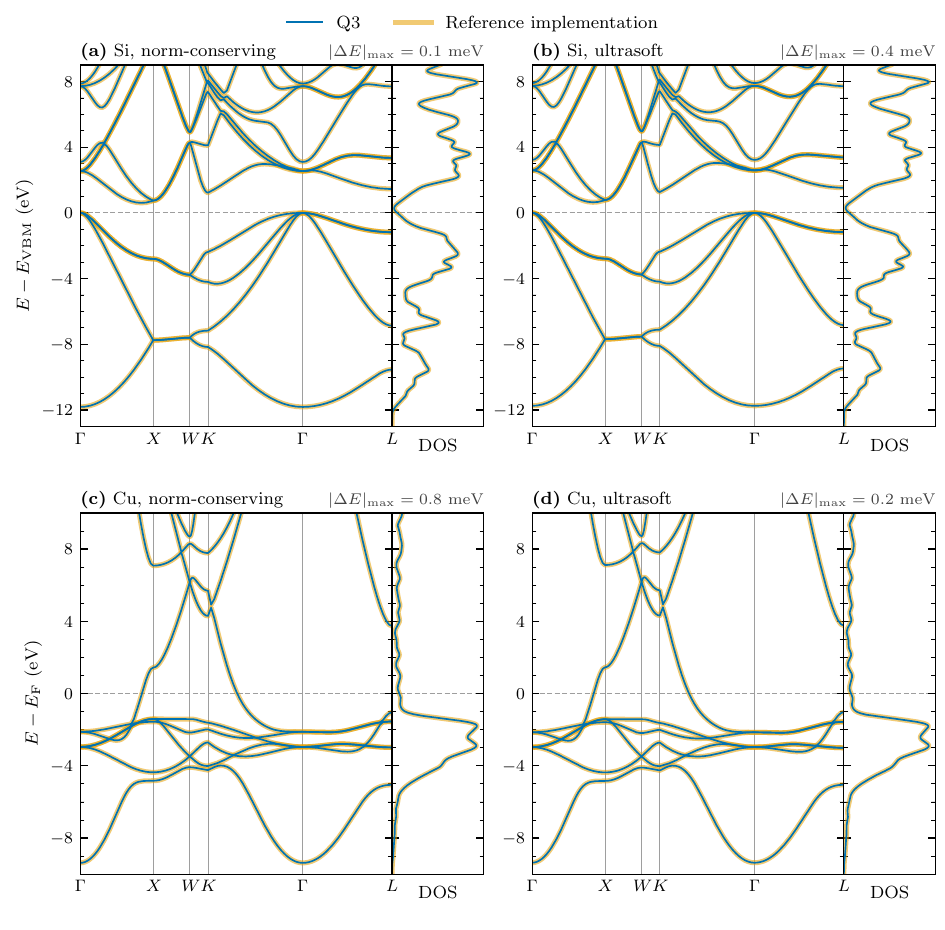}
            \caption{Band structures along $\Gamma$--$X$--$W$--$K$--$\Gamma$--$L$ and densities of states (DOS, right side of each panel) computed with Q3 (thin blue lines) and the reference implementation \cite{QE2009mainReference} (wide orange lines) for Si with (a) norm-conserving and (b) ultrasoft pseudopotentials, and Cu with (c) norm-conserving and (d) ultrasoft pseudopotentials. Both codes use identical structures, pseudopotentials, cutoffs, k-point grids, and the LibXC PBE functional. Energies are relative to the valence-band maximum for Si and to the Fermi level for Cu (dashed lines). The DOS is obtained from the self-consistent eigenvalues with a Gaussian broadening of 0.25~eV. The maximum absolute eigenvalue deviation along the path, $|\Delta E|_{\max}$, is given above each panel.}
            \label{fig:band-structures}
        \end{figure}

        Figure~\ref{fig:band-structures} compares the band structures and densities of states of silicon (a, b) and copper (c, d) computed with norm-conserving (a, c) and ultrasoft (b, d) pseudopotentials. The Q3 eigenvalues (thin lines) lie on top of those of the reference implementation (wide lines) along the entire high-symmetry path, including the narrow copper d-band manifold: the maximum deviation over all reported bands and k-points is below 1~meV in every case.

        In addition to NC and US validation, Q3's PAW implementation is validated end-to-end using redistributable UPF v2 datasets across ground-state SCF, spin polarization (LSDA), Hellmann-Feynman forces, Nielsen-Martin stress tensors, and Dudarev DFT+U. Against a LibXC-enabled Quantum ESPRESSO build, PAW total energies agree to within 1~mRy/atom across silicon, copper, and diamond. The small residual arises from different one-center radial discretizations inside the PAW augmentation spheres; as a per-element constant offset, it cancels in energy differences, forces, and stresses. PAW forces and stress tensors are gated directly against numerical finite-difference derivatives of Q3's own total energy: on silicon PAW, the isotropic pressure matches the fixed-basis volume derivative, stress is invariant under origin shifts, and $\Gamma$-only forces match $\mathrm{d}E/\mathrm{d}\mathbf{R}$ to within $9 \times 10^{-5}$ relative error; on zincblende SiC with atomic and ortho-atomic DFT+U projectors, forces agree with $\mathrm{d}E/\mathrm{d}\mathbf{R}$ to $2.4 \times 10^{-6}$~Ry/Bohr ($0.013\%$ relative) and all six stress tensor components agree within $0.05$~kbar. Cross-code DFT+U calculations reading Quantum ESPRESSO's own ultrasoft Hubbard input for silicon match total energies to $4.5 \times 10^{-7}$~Ry ($3 \times 10^{-8}$~Ry for the Hubbard energy $E_U$) with atomic projectors and $5.8 \times 10^{-7}$~Ry ($1.1 \times 10^{-8}$~Ry for $E_U$) with ortho-atomic projectors.

        \subsubsection{Surfaces and Interfaces}
        \label{subsubsec:results-accuracy-surfaces}

        \begin{figure}[H]
            \centering
            \includegraphics[width=0.6\textwidth]{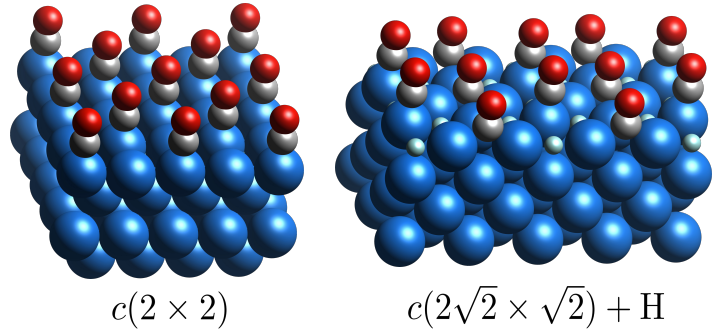}
            \caption{Slab geometries of CO chemisorbed in c$(2 \times 2)$ on-top configuration (left) and co-adsorbed with H in hollow sites in c$(2\sqrt{2} \times \sqrt{2})$ configuration (right), where half the CO molecules bind to hollow and the other half to top sites. Ni, C, O, and H are represented by blue, gray, red, and cyan spheres, respectively.}
            \label{fig:co_ni100}
        \end{figure}

        Extending beyond the bulk magnetic systems, we further compare Q3 against Quantum ESPRESSO for magnetic surfaces with chemisorbed adsorbates. We consider CO chemisorbed on Ni(100) at different coverages with and without H as a coadsorbate (see Figure~\ref{fig:co_ni100}). Coadsorption and surface coverage allow for experimental preparation of CO bound to different high-symmetry sites on Ni(100); when probed with C- and O-K core-edge X-ray spectroscopy, these different binding sites lead to pronounced spectral shifts due to changes in the core-hole binding energy of the core-excited atom \cite{WesterlundSurfSci1988}.

        Frozen-core hole pseudopotential DFT, where the core-excitation is modeled with custom pseudopotentials, was shown previously to reproduce these X-ray spectral shifts on Ni(100) \cite{DiesenAIPAdv2020} with Ni slabs constructed by employing experimental lattice constants. Here, to establish a further benchmark between Q3 and Quantum ESPRESSO, we have instead performed equation of state fits based on Q3 and Quantum ESPRESSO elastic curves first, respectively. At a planewave cutoff of 520~eV, a $k$-point density of $\sim$0.02~\AA$^{-1}$, and employing the RPBE XC functional \cite{HammerPRB1999} with an ultrasoft pseudopotential, a lattice constant of 3.558~{\AA} was found, with a difference between Q3 and Quantum ESPRESSO of only $3 \cdot 10^{-6}$~\AA, which is a negligible difference for lattice constants. We use this lattice constant to construct slabs, where the bottom two layers are fixed at bulk coordinates, and the top two layers and adsorbates are relaxed.

        \begin{table}[H]
            \centering
            \MateraTableSetup
            \caption{Core-hole binding energy shifts for the K-edges of C and O for different adsorbate configurations of CO on Ni(100) computed with Q3 and Quantum ESPRESSO. Experimental shifts derived from X-ray photoelectron spectroscopy from Ref.~\cite{WesterlundSurfSci1988}. All shifts in eV and relative to c$(2 \times 2)$. Theoretical shifts shown with meV resolution to highlight the agreement between Q3 and Quantum ESPRESSO (the results also agree relatively well with previous theoretical studies \cite{DiesenAIPAdv2020} performed at experimental rather than relaxed lattice constant). Agreement between theory and experiment is sufficient to use DFT as a discriminator of binding sites from observed X-ray spectral shifts.}
            \label{tab:xpsshifts}
            \begin{tabularx}{\linewidth}{|l|>{\centering\arraybackslash}X|>{\centering\arraybackslash}X|>{\centering\arraybackslash}X|>{\centering\arraybackslash}X|>{\centering\arraybackslash}X|}
                \hline
                \MateraTH{Config.} & \MateraTH{c$(2 \times 2)$} & \MateraTH{c$(2 \times 2) + \textrm{H}$} & \MateraTH{p$(3\sqrt{2} \times \sqrt{2})$} & \MateraTH{c$(2\sqrt{2} \times \sqrt{2})$ $+ \textrm{H} \,(^*\textrm{top})$} & \MateraTH{c$(2\sqrt{2} \times \sqrt{2})$ $+ \textrm{H} \,(^*\textrm{hollow})$} \\ \hline\hline
                C exp. & $0$ & $0.4$   & $-0.4$   & $0.3$   & $-0.8$   \\ \hline
                C Q3   & $0$ & $0.304$ & $-0.425$ & $0.233$ & $-0.547$ \\ \hline
                C QE   & $0$ & $0.304$ & $-0.425$ & $0.233$ & $-0.547$ \\ \hline\hline
                O exp. & $0$ & $0.7$   & $-0.9$   & $0.6$   & $-1.9$   \\ \hline
                O Q3   & $0$ & $0.543$ & $-1.038$ & $0.431$ & $-1.836$ \\ \hline
                O QE   & $0$ & $0.543$ & $-1.038$ & $0.431$ & $-1.836$ \\ \hline
            \end{tabularx}
        \end{table}

        The core-excitation process is then modeled by replacing one C or one O pseudopotential with a special pseudopotential, where one 1$s$ electron is missing in the core. As total energies computed with different pseudopotentials can generally not be compared, this approach does not provide absolute core-hole binding energies, but shifts in binding energies can be computed. Here, the considered shifts are due to different binding sites of CO. The computed shifts (see Table~\ref{tab:xpsshifts}) due to different binding sites differ between Q3 and Quantum ESPRESSO by only {$\sim$}$10^{-5}$~eV, a negligible difference on this energy scale. This is a demonstration of Q3 being able to use different types of pseudopotentials including custom frozen-core hole types.

        \subsubsection{Beyond DFT: Many-Body Workflows (Q3 + BerkeleyGW) example}
        \label{subsubsec:results-accuracy-beyond-dft}

        To demonstrate Q3's readiness as a mean-field driver for downstream many-body perturbation theory (GW/BSE) calculations, we validated the native \texttt{q3-util convert} post-processing pipeline on bulk diamond-cubic silicon using a norm-conserving pseudopotential. Starting from a self-consistent Q3 ground-state calculation on a symmetry-reduced k-point mesh, \texttt{q3-util convert} generates the complete set of binary and ASCII interface files required by BerkeleyGW \cite{deslippe2012berkeleygw}: the plane-wave wavefunction file (\texttt{WFN}), the reciprocal-space charge density (\texttt{RHO}), and the diagonal and off-diagonal exchange-correlation matrix elements $\langle n\mathbf{k} | V_{\text{xc}} | m\mathbf{k} \rangle$ (\texttt{vxc.dat}). The exported wavefunctions satisfy unitarity and orthonormality across all exported bands and k-points to machine precision ($< 10^{-14}$), the reciprocal-space charge density sum rule reproduces the valence electron count ($N_e = 8.0$) to within $10^{-12}$, and the exported $\langle n\mathbf{k} | V_{\text{xc}} | m\mathbf{k} \rangle$ matrix elements agree with Q3's internal Kohn-Sham operator projections to within $2.3 \times 10^{-14}$~eV, confirming lossless coupling between Q3 and external many-body workflows.

    \subsection{Single-Core Performance and Algorithmic Optimizations}
    \label{subsec:results-performance}

        Single-core wall-clock timings (Apple M2) show that Q3 is consistently faster than the reference implementation across a range of system sizes and physical scenarios. Both codes (Q3 and Quantum ESPRESSO 7.5, \texttt{pw.x}) ran single-threaded with identical pseudopotentials, plane-wave cutoffs, $\Gamma$-only sampling, occupations, and an SCF convergence threshold of $10^{-8}$~Ry. Table~\ref{tab:supercell-benchmarks} presents the results for $3\times3\times3$ supercell SCF calculations (27--54 atoms) at equilibrium geometries, while Table~\ref{tab:displaced-benchmarks} presents results for thermally displaced geometries ($\sim$0.02--0.04~Bohr random displacements), which more closely mimic real-world use cases such as molecular dynamics or phonon calculations. Q3 achieves speedups of 1.2--5.3$\times$ compared to the reference code, with the largest gains observed for norm-conserving copper supercells under displaced geometries. At equilibrium, both codes need a similar number of SCF iterations, and in relaxations Q3 takes the same number of BFGS steps or one more, so the speedups mainly reflect a lower cost per step rather than fewer steps. For ionic relaxations (Table~\ref{tab:relaxation-benchmarks}), Q3 is driven by ASE's BFGS through its in-memory Python bindings and the reference code by its native BFGS, so both reuse wavefunctions between ionic steps.

        \begin{table}[H]
            \centering
            \MateraTableSetup
            \caption{Single-core SCF wall time, $3\times3\times3$ supercells at equilibrium geometry (Q3 vs.\ Reference \cite{QE2009mainReference}).}
            \label{tab:supercell-benchmarks}
            \begin{tabularx}{\linewidth}{|l|l|c|Q|Q|c|}
                \hline
                \MateraTH{System} & \MateraTH{PP} & \MateraTH{Atoms} & \MateraTH{Q3 (s)} & \MateraTH{Reference \cite{QE2009mainReference} (s)} & \MateraTH{Speedup} \\
                \hline
                Si & NC & 54 & 39.0 & 107.7 & 2.8$\times$ \\ \hline
                Si & US & 54 & 40.8 & 94.1  & 2.3$\times$ \\ \hline
                Cu & NC & 27 & 52.3 & 122.9 & 2.3$\times$ \\ \hline
                Cu & US & 27 & 28.4 & 34.5  & 1.2$\times$ \\ \hline
            \end{tabularx}
        \end{table}

        \begin{table}[H]
            \centering
            \MateraTableSetup
            \caption{Single-core SCF wall time, $3\times3\times3$ supercells at thermally displaced geometry (Q3 vs.\ Reference \cite{QE2009mainReference}).}
            \label{tab:displaced-benchmarks}
            \begin{tabularx}{\linewidth}{|l|l|c|Q|Q|c|}
                \hline
                \MateraTH{System} & \MateraTH{PP} & \MateraTH{Atoms} & \MateraTH{Q3 (s)} & \MateraTH{Reference \cite{QE2009mainReference} (s)} & \MateraTH{Speedup} \\
                \hline
                Si & NC & 54 & 39.3 & 125.7 & 3.2$\times$ \\ \hline
                Si & US & 54 & 41.8 & 86.9  & 2.1$\times$ \\ \hline
                Cu & NC & 27 & 81.0 & 429.7 & 5.3$\times$ \\ \hline
                Cu & US & 27 & 40.8 & 87.9  & 2.2$\times$ \\ \hline
            \end{tabularx}
        \end{table}

        \begin{table}[H]
            \centering
            \MateraTableSetup
            \caption{Ionic relaxation wall time, $3\times3\times3$ supercells from displaced geometry (Q3 with ASE BFGS vs.\ Reference \cite{QE2009mainReference} with its native BFGS).}
            \label{tab:relaxation-benchmarks}
            \begin{tabularx}{\linewidth}{|l|l|c|Q|Q|c|}
                \hline
                \MateraTH{System} & \MateraTH{PP} & \MateraTH{Atoms} & \MateraTH{Q3 (s)} & \MateraTH{Reference \cite{QE2009mainReference} (s)} & \MateraTH{Speedup} \\
                \hline
                Si & NC & 54 & 145.6 & 335.4 & 2.3$\times$ \\ \hline
                Si & US & 54 & 245.7 & 388.8 & 1.6$\times$ \\ \hline
                Cu & NC & 27 & 589.2 & 1085.5 & 1.8$\times$ \\ \hline
                Cu & US & 27 & 146.4 & 348.1 & 2.4$\times$ \\ \hline
            \end{tabularx}
        \end{table}

        Q3's single-core performance rests on several implementation choices, most of which follow established practice in mature plane-wave codes:
        \begin{itemize}
            \item \textbf{BLAS 3 Block Orthogonalization}: The Davidson diagonalizer replaces band-by-band Gram-Schmidt projections with block operations utilizing high-level BLAS 3 routines (\texttt{dgemm}/\texttt{zgemm}) for projecting out the existing basis, and a fast Cholesky QR algorithm (\texttt{dpotrf}/\texttt{dtrsm}) for intra-batch orthogonalization. Within Q3, this reduces orthogonalization time by up to $16\times$ for large active subspaces.
            \item \textbf{Smooth Reciprocal-Space Mixing}: For systems with ultrasoft pseudopotentials, the Broyden density mixer restricts its reciprocal-space history storage and vector operations strictly to the smooth G-vector grid. By excluding the highly oscillatory ultrasoft augmentation charge components from the Broyden subspace, the mixer converges in fewer steps and avoids numerical instability. For the displaced copper supercells, Q3 also used a deeper Broyden history than the reference run (32 vs.\ 8 vectors), which lowers its SCF iteration count.
            \item \textbf{FFT Buffer Reuse and Zero-Copy Operations}: Q3 pre-allocates and reuses internal buffers for FFT grids and Davidson workspaces to eliminate runtime allocation overhead. Furthermore, the Hamiltonian application uses a zero-copy FFT interface that scatters and gathers G-space wavefunction coefficients directly to and from internal FFTW3 buffers.
            \item \textbf{In-Memory Caching and Restarts}: Utilizing the stateful C++ library API via in-memory Python bindings (\texttt{q3\_core}), geometry relaxation workflows cache pseudopotential sets and preload the wavefunctions and charge densities from the previous ionic step. This matches the warm-start behavior of the reference code's native relaxation loop.
        \end{itemize}

    \subsection{MPI K-Point Pool Parallel Scaling}
    \label{subsec:results-mpi}

        Q3 implements MPI k-point pool parallelism, distributing pools of k-points across independent MPI processes. Table~\ref{tab:mpi-scaling} presents parallel scaling results for a copper primitive cell with an $8\times8\times8$ k-point grid (60 irreducible k-points after symmetry reduction) using a norm-conserving pseudopotential at $E_{\text{wfc}} = 120$~Ry.

        \begin{table}[H]
            \centering
            \MateraTableSetup
            \caption{MPI k-point pool parallel scaling for Copper NC ($8\times8\times8$ k-grid, 60 irreducible k-points; Q3 vs.\ Reference \cite{QE2009mainReference}).}
            \label{tab:mpi-scaling}
            \begin{tabularx}{\linewidth}{|c|Q|c|c|Q|c|c|}
                \hline
                \MateraTH{MPI Ranks} & \MateraTH{Q3 (s)} & \MateraTH{Q3 Speedup} & \MateraTH{Q3 Eff.} & \MateraTH{Ref \cite{QE2009mainReference} (s)} & \MateraTH{Ref Speedup} & \MateraTH{Q3 / Ref} \\
                \hline
                1 & 9.47 & 1.00$\times$ & 100\% & 16.84 & 1.00$\times$ & \textbf{0.56$\times$} \\ \hline
                2 & 5.35 & 1.77$\times$ & 88\%  & 8.51  & 1.98$\times$ & \textbf{0.63$\times$} \\ \hline
                4 & 3.70 & 2.56$\times$ & 64\%  & 5.55  & 3.03$\times$ & \textbf{0.67$\times$} \\ \hline
            \end{tabularx}
        \end{table}

        Key observations from the MPI scaling benchmarks include:
        \begin{itemize}
            \item \textbf{Consistently Faster Across Ranks}: Q3 is $\sim 1.8\times$ faster than the reference code at 1 rank (0.56$\times$ runtime) and maintains a speedup advantage across all rank counts.
            \item \textbf{Numerical Topology Invariance}: Total energies across 1, 2, and 4 MPI ranks match to within $10^{-9}$~Ry, confirming that inter-pool MPI reduction routines for charge density, Fermi energy bisection, and total energy components are topology-invariant to full SCF precision.
            \item \textbf{Scaling Scope}: These benchmarks demonstrate correctness of the k-point pool parallelization framework on a shared-memory workstation, where memory bandwidth saturation limits parallel efficiency beyond 2 ranks. Measurements on a leadership HPC system are reported below.
        \end{itemize}

        Table~\ref{tab:perlmutter-scaling} summarizes distributed scaling measured on NERSC Perlmutter \cite{nersc2021perlmutter} (CPU nodes: dual AMD EPYC 7763; GPU nodes: NVIDIA A100-SXM4-40GB). Times are the solver's own reported totals, and the total energy of every run is checked against the serial result.

        \begin{table}[H]
            \centering
            \MateraTableSetup
            \caption{Distributed scaling on NERSC Perlmutter \cite{nersc2021perlmutter}. Si: primitive cell, $8\times8\times8$ k-grid (60 irreducible k-points), NC. Graphite: $12\times12\times6$ k-grid (76 irreducible k-points), NC with DFT-D3, capped at three SCF iterations. One k-point pool per rank (CPU) or per GPU.}
            \label{tab:perlmutter-scaling}
            \begin{tabularx}{\linewidth}{|l|l|Q|c|c|}
                \hline
                \MateraTH{System} & \MateraTH{Executable} & \MateraTH{Resources} & \MateraTH{Nodes} & \MateraTH{Speedup} \\
                \hline
                Si       & \texttt{q3\_mpi}          & 8 ranks  & 1 & 7.55$\times$ \\ \hline
                Si       & \texttt{q3\_mpi}          & 32 ranks & 1 & 28.9$\times$ \\ \hline
                Graphite & \texttt{q3\_mpi}          & 16 ranks & 1 & 20.4$\times$ \\ \hline
                Graphite & \texttt{q3\_mpi}          & 16 ranks & 2 & 21.9$\times$ \\ \hline
                Si       & \texttt{q3\_mpi\_gpu\_cuda} & 4 A100   & 1 & 4.19$\times$ \\ \hline
                Si       & \texttt{q3\_mpi\_gpu\_cuda} & 8 A100   & 2 & 7.88$\times$ \\ \hline
            \end{tabularx}
            \vspace{2mm}
            {\small\textit{Note}: CPU speedups are relative to one CPU rank; GPU speedups are relative to one A100 running the same executable. The 16-rank graphite efficiency above 100\% reflects cache effects and the higher run-to-run spread of its serial baseline.}
        \end{table}

        Three results stand out. First, k-point pools scale near-linearly up to the irreducible k-point count (90\% efficiency at 32 ranks for silicon), and crossing the node boundary costs nothing measurable: at matched rank counts the two-node graphite runs are as fast as the single-node runs, and all twenty graphite runs agree to $1.4\times10^{-14}$~Ry. Second, the same pool decomposition carries over to GPUs, with one device per MPI rank reaching 98\% efficiency on eight A100s across two nodes and energies identical to the single-GPU run. Third, intra-pool parallelism is primarily a memory lever: distributing reciprocal space by FFT stick reduces the peak resident memory per rank for a 54-atom silicon supercell from 1.60~GiB at two ranks to 0.70~GiB at eight (measured on a 10-core Apple M-series workstation), with total energies unchanged to $10^{-9}$~Ry, whereas on Perlmutter the speed gain is modest with the default reciprocal-space split ($1.13\times$ on four ranks for a $\Gamma$-only Cu $3\times3\times3$ supercell) and larger with the band-distribution mode ($2.59\times$ on the same four ranks).

    \subsection{GPU Acceleration}
    \label{subsec:results-gpu}

        To support next-generation accelerated computing, Q3 implements a modular GPU execution layer designed for NVIDIA CUDA and Apple Metal backends. The implementation features a GPU-resident Davidson solver workspace (\texttt{GPUDavidsonWorkspace}) that keeps wavefunctions and Ritz vectors entirely on the device, eliminating PCIe transfer overhead. Performance-critical operations are offloaded to GPU-accelerated libraries (cuBLAS, cuFFT, cuSOLVER), with real symmetric projections and subspace updates executed via double-precision matrix multiplications on the device.

        Table~\ref{tab:gpu-benchmarks} presents benchmarks for a Copper $3\times3\times3$ supercell calculation (27 atoms, 280 bands, 192,228 G-vectors, $\Gamma$-point, 120~Ry cutoff) on an NVIDIA Tesla T4 GPU (16~GB GDDR6, Turing architecture, CUDA 12.8). While the T4 is an entry-level datacenter GPU (with limited FP64 throughput at 1/32 of FP32 peak), it serves as a controlled comparison that attributes the gain to the GPU-resident solver rather than to the hardware. It is not a measure of what a GPU is worth against a well-configured CPU node: on Perlmutter, the same system runs $4.1\times$ faster on an A100 than on one CPU core, but at roughly parity with 8 CPU threads, because 280 bands do not occupy the device (Section~\ref{subsec:discussion-performance}). The GPU-accelerated solver achieves a $6.5\times$ speedup in full double precision (FP64) and a $6.6\times$ overall speedup with mixed-precision execution, with the SCF iteration kernel itself accelerated by $10.8\times$.

        \begin{table}[H]
            \centering
            \MateraTableSetup
            \caption{Q3 GPU performance (NVIDIA Tesla T4 vs single-core CPU, Cu NC $3\times3\times3$, 280 bands, $\Gamma$-only).}
            \label{tab:gpu-benchmarks}
            \begin{tabularx}{\linewidth}{|l|Q|c|Q|c|}
                \hline
                \MateraTH{Execution Mode} & \MateraTH{Total SCF (s)} & \MateraTH{Speedup} & \MateraTH{SCF Iteration (s)} & \MateraTH{Iter. Speedup} \\
                \hline
                Serial (CPU) & 411.0 & 1.0$\times$ & 384.0 & 1.0$\times$ \\ \hline
                CUDA FP64 & 63.6 & 6.5$\times$ & 36.7 & 10.5$\times$ \\ \hline
                CUDA FP32 (Mixed) & 62.3 & 6.6$\times$ & 35.3 & 10.8$\times$ \\ \hline
            \end{tabularx}
            \vspace{2mm}
            {\small\textit{Note}: The converged total energy difference between the CUDA FP64 and CUDA FP32 (Mixed) runs is $< 10^{-7}$~Ry ($< 0.1$~$\mu$Ry), confirming that mixed-precision execution maintains full double-precision accuracy in the final result.}
        \end{table}

        Key GPU optimizations include:
        \begin{itemize}
            \item \textbf{GPU-Resident Workspace}: Wavefunctions, Ritz vectors, and subspace matrices remain GPU-resident throughout the Davidson iteration, achieving zero PCIe page-fault transfer overhead.
            \item \textbf{Orthogonalization Offload}: Gram-Schmidt projections via \texttt{dgemm} on GPU reduce orthogonalization time from 90.6~s to 9.96~s ($9\times$ speedup).
            \item \textbf{Subspace Update Offload}: Subspace rotation and update operations are reduced from 11.4~s to 1.01~s ($11\times$ speedup) on device.
            \item \textbf{Mixed-Precision Policy}: FP32 wavefunction FFTs with FP64 energy accumulation maintain sub-$\mu$Ry accuracy while providing additional acceleration.
        \end{itemize}

    \subsection{Developer Velocity}
    \label{subsec:results-velocity}

        Perhaps the most significant result is the development velocity enabled by the AI-native architecture. The core Q3 codebase, comprising $\sim$82,200 lines of C++ across 461 header files and 29 translation units together with over 1,800 automated tests (1,637 C++ unit/integration tests, 218 Python validation tests, and 32 regression tests), was developed in a matter of weeks of active development utilizing frontier agentic AI coding tools (including Anthropic Claude Code, Google Gemini / Antigravity, and Cursor) configured with strict project-level architectural rules (\texttt{AGENTS.md}). In this workflow, human domain scientists defined the layered object-oriented interfaces, physical equations, and numerical tolerance targets, while AI agents generated the class implementations, unit tests, and build configurations within isolated worktrees gated by automated CI. Rather than deferring verification to the end of the project, every physical capability, progressing from core types, symmetry, and norm-conserving SCF to ultrasoft and PAW pseudopotentials, forces, stress tensors, magnetism, DFT+U, dispersion, multi-vendor GPU backends, and many-body exports, was paired immediately with analytical unit tests and cross-code validation benchmarks.

        We intentionally refrain from quoting specific human-month or headcount figures at this stage, as human-supervised agentic development interleaves autonomous code and test synthesis with human architectural steering and physical validation in ways that do not map directly onto conventional person-month software engineering metrics. Even so, compressing the creation of a production-grade, multi-GPU plane-wave DFT engine from the multiple person-years typically required for established codes into a matter of weeks demonstrates that Q3's AI-native design principles fundamentally lower the barrier to developing and extending scientific software.

    \section{Discussion}
\label{sec:discussion}

    Quantum 3xplorer (Q3) represents a shift in how scientific simulation software is designed, developed, and maintained. Building on the theoretical and numerical foundations established by community electronic-structure codes, Q3 introduces an AI-native software architecture tailored for human-AI co-development and modern data-driven workflows.

    \subsection{The Impact of AI-Native Design and Comparison with Modern Frameworks}
    \label{subsec:discussion-ai-impact}

        The primary motivation behind Q3 is the realization that accelerating materials discovery increasingly requires modernizing the software engineering workflow alongside physical theory. Established Fortran-based engines remain remarkably fast and scientifically comprehensive, yet their procedural, shared-state architectures were not designed for automated refactoring or reasoning by large language models (LLMs). Q3's commitment to explicit naming, modular layering, and structured data formats makes the engine a first-class citizen in the era of AI-assisted coding. Crucially, the one-class-per-file convention and the strict avoidance of domain-specific abbreviations reduce the context-window overhead for LLMs, enabling AI coding agents to reason about individual components without loading entire subsystem codebases. The rapid development of validated band-structure, forces, stress, spin-polarized, PAW, and dispersion-corrected capabilities in a matter of weeks of active development highlights the productivity gains possible when a codebase is architected for transparency.

        It is instructive to situate Q3 alongside two prominent recent efforts that also rethink plane-wave DFT software design: \textbf{DFTK} \cite{herbst2021dftk} and \textbf{SIRIUS} \cite{zhang2023sirius} (Table~\ref{tab:modern-dft-comparison}). DFTK is written in pure Julia ($\sim$10,000 lines of code) to bridge applied mathematics and electronic-structure theory, emphasizing concise mathematical expressiveness and automatic differentiation in a high-level dynamic language. SIRIUS is a domain-specific C++14/17 electronic-structure code with high-performance GPU and MPI support that runs standalone and also serves as an execution backend accelerating inner-loop kernels within host Fortran codes such as Quantum ESPRESSO, CP2K, and exciting. What distinguishes Q3 is not merely its language choice or feature set, but its \textbf{AI-native concept}: an end-to-end simulation engine built to be \textit{understandable for AI, written by AI, maintained by AI, and reviewed and validated by humans}. Rather than optimizing strictly for human mathematical brevity (DFTK) or host-coupled kernel acceleration (SIRIUS), Q3 aligns its entire architecture so that AI agents can autonomously author, extend, and audit production C++/GPU code under human scientific supervision.

        \begin{table}[H]
            \centering
            \MateraTableSetup
            \caption{Conceptual comparison of Q3 with representative established and modern plane-wave DFT frameworks.}
            \label{tab:modern-dft-comparison}
            \begin{tabularx}{\linewidth}{|l|l|X|X|}
                \hline
                \MateraTH{Framework} & \MateraTH{Primary Language} & \MateraTH{Core Design Paradigm} & \MateraTH{Relationship to Q3} \\
                \hline
                Quantum ESPRESSO \cite{QE2009mainReference} & Fortran 90/2008 & Comprehensive community suite for plane-wave DFT, DFPT, and excitations & Primary numerical validation reference and input/pseudopotential standard \\ \hline
                SIRIUS \cite{zhang2023sirius} & C++14/17 + CUDA/ROCm & Domain-specific HPC electronic-structure code, also used to accelerate DFT kernels inside host Fortran codes & Shares C++/GPU performance goals; Q3 is a standalone AI-native engine rather than a host-code backend \\ \hline
                DFTK \cite{herbst2021dftk} & Julia & Concise mathematical workbench for algorithmic research and automatic differentiation & Shares differentiability and legibility goals; Q3 targets compiled C++17 production workflows \\ \hline
                \textbf{Quantum 3xplorer (Q3)} & \textbf{C++17 + Python} & \textbf{AI-native lifecycle: understandable for AI, written by AI, maintained by AI, reviewed/validated by humans} & \textbf{Standalone engine introducing human-supervised, AI-authored scientific software development} \\ \hline
            \end{tabularx}
        \end{table}

    \subsection{Ecosystem Integration and Community-driven Data Standards}
    \label{subsec:discussion-ecosystem}

        A core design principle of Q3 is its strict compliance with open data and representation standards. Rather than relying on custom text outputs that require fragile regex parsing, Q3 leverages the CAMSAI standards \cite{camsai-standards, mat3ra-esse}. The engine's inputs and outputs are validated directly against structured JSON schemas, which are shared with other compatible tools and platforms.
        
        This integration allows Q3 to be natively driven by compatible tools and platforms, such as structure generators and workflow managers. By standardizing physical properties, crystal structures, and convergence parameters into machine-readable formats, Q3 bridges the gap between raw DFT simulations and automated database curation. This data model compatibility ensures that generated datasets are immediately searchable, inspectable, and reusable in high-throughput workflows. Additionally, Q3 provides built-in drop-in compatibility with Quantum ESPRESSO input files via a native Fortran namelist parser, while its build-time add-on registries allow optional modules to support workflows from other community or proprietary simulation packages without manual reformatting or coupling the core engine to external conventions.

    \subsection{Toward a Unified AI-Native Quantum Simulation Engine}
    \label{subsec:discussion-unified-engine}

        A central long-term vision for Q3 is to serve as a single, unified, AI-native engine capable of spanning the full hierarchy of quantum simulation methods. The current implementation establishes a production-quality DFT foundation, but the strictly layered architecture, with its clean separation of abstract quantum-mechanical interfaces (Layer~2) from concrete representations (Layer~3), is deliberately designed to accommodate methods far beyond mean-field theory.

        The method-agnostic core enables the progressive integration of many-body perturbation theory (GW and BSE calculations \cite{hybertsen1986gw, rohlfing2000bse, deslippe2012berkeleygw}), quantum Monte Carlo methods, and advanced correlated-electron approaches within the same codebase. Rather than requiring separate, disconnected software packages for each level of theory, Q3 provides a unified framework where different methods share the same data structures, I/O infrastructure, and hardware acceleration layers. The existing BerkeleyGW mean-field export pipeline (\texttt{q3-util convert}) already demonstrates this interoperability: converged DFT wavefunctions flow directly into quasiparticle and optical property calculations without format conversion or data loss. A native Wannier interpolation engine \cite{marzari2012wannier}, constructing projection-gauge Wannier functions from overlap and projection matrices computed by Q3 itself and validated against direct Q3 band calculations as well as the interface matrices of an established plane-wave code's Wannier90 \cite{pizzi2020wannier90} interface (0.6~meV rms along $\Gamma$--X--L--$\Gamma$ for silicon), enables Brillouin zone interpolation from a coarse DFT mesh to arbitrary k-point paths and dense integration grids for isolated band manifolds.

        Looking further ahead, the engine's architecture is designed to interface with quantum computing backends, where hybrid classical-quantum algorithms (such as the Variational Quantum Eigensolver) could leverage Q3's mean-field solutions as initial states. By consolidating these diverse simulation paradigms into a single, legible, and AI-extensible codebase, Q3 aims to eliminate the fragmentation that currently forces researchers to maintain expertise across multiple incompatible software ecosystems.

    \subsection{Extensibility and Differentiable Simulation}
    \label{subsec:discussion-differentiable}

        A critical feature of Q3's design is its method-agnostic core. By separating abstract quantum-mechanical interfaces from the concrete plane-wave representation, Q3 serves as a platform for diverse simulation methods. The architectural roadmap includes the implementation of a differentiable interface that returns exact gradients (forces, stress, and electronic derivatives) alongside energies, which is critical for machine-learned interatomic potentials like NequIP \cite{batzner2022nequip} and MACE \cite{batatia2022mace}.

        Q3's layered design is particularly well-suited for differentiable simulation. The strict separation of concerns means that the Kohn-Sham self-consistency loop can be treated as a fixed-point iteration, enabling implicit differentiation without backpropagating through the full SCF trajectory. The existing Broyden and Anderson mixing infrastructure can be repurposed as the linear solver for the implicit differentiation system, while the Hellmann-Feynman theorem provides exact position gradients for norm-conserving plane-wave bases without basis-set Pulay corrections (for ultrasoft pseudopotentials, the additional augmentation charge derivatives are computed analytically). The autodiff framework is designed to trace through Ewald summation \cite{ewald1921}, pseudopotential projectors, and cell-vector degrees of freedom. This architectural readiness, combined with planned PyTorch \texttt{autograd.Function} and JAX \texttt{ffi} bindings with DLPack zero-copy GPU memory sharing, positions Q3 as an active element in deep learning pipelines rather than a passive data-generation utility \cite{kasim2021diffdft, li2021ksregularizer}.

    \subsection{Performance, Portability, and HPC Deployment}
    \label{subsec:discussion-performance}

        With MPI k-point pool and intra-pool parallelism and GPU acceleration backends implemented across four vendor targets, Q3 has been deployed and benchmarked on NERSC's Perlmutter supercomputer \cite{nersc2021perlmutter}. MPI k-point pool parallelism scales near-linearly, achieving $28.9\times$ speedup on 32 ranks (90\% parallel efficiency) for a silicon primitive cell with 60 irreducible k-points, and crosses the node boundary without measurable cost (Section~\ref{subsec:results-mpi}). The multi-GPU executable assigns one A100 per MPI rank and one k-point pool per device, reaching $7.88\times$ on eight GPUs across two nodes (98\% efficiency) with energies identical to the single-GPU run. On a single device, the A100 delivers $4.1\times$ over one CPU core in double precision with numerically identical energy, but only $1.37\times$ over 32 OpenMP threads on the same node and roughly parity with the best CPU configuration measured (8 threads), as the $\Gamma$-only 280-band test system is too small to occupy the device. The vendor-agnostic GPU abstraction layer has been validated on NVIDIA CUDA and Apple Metal (with numerically identical CPU parity), while AMD HIP and Intel SYCL backends compile and run in CI, awaiting production hardware validation.

        Additionally, Q3's clean electronic-structure outputs can serve as high-quality starting points for many-body perturbation theory calculations. For instance, quasiparticle and optical property calculations utilizing GW/BSE methods, implemented in codes like BerkeleyGW \cite{deslippe2012berkeleygw}, require highly converged and numerically stable mean-field wavefunctions and eigenvalues. The standalone \texttt{q3-util convert} subcommand exports converged mean-field records directly to BerkeleyGW formats (\texttt{WFN}, \texttt{RHO}, \texttt{vxc.dat}), enabling seamless post-DFT workflows without re-running SCF calculations.

    \subsection{Code Provenance and Licensing Integrity}
    \label{subsec:discussion-provenance}

        Releasing an AI-native scientific simulation engine under the permissive Apache-2.0 license requires rigorous assurances of code provenance, particularly in a domain where established reference implementations carry strong copyleft (e.g., GPLv2-or-later) or proprietary commercial licenses. While AI coding agents dramatically accelerate software development, they also require strict guardrails to ensure that generated implementations are derived from primary mathematical literature rather than translated or paraphrased from incompatibly licensed source code. In Q3, algorithms and physical reference tables are implemented from scratch from published equations, format specifications, and standard constants, while cross-code validation against established engines is performed strictly on black-box numerical outputs (energies, forces, stresses, and eigenvalues). To enforce and continuously verify this provenance, Q3 employs a three-layered assurance framework:

        \begin{itemize}
            \item \textbf{Automated Checks in the Codebase and CI Pipeline}: Every commit requires a Developer Certificate of Origin (\texttt{Signed-off-by} DCO trailer) enforced both locally at commit time and in continuous integration, paired with tree-wide, machine-readable copyright and license declarations maintained in a \texttt{REUSE.toml} manifest. Pre-commit hooks and CI linters enforce Q3's structural and stylistic conventions (including downward-only C++ include and Python import layering, one-class-per-file organization, ratchets against dynamic type-checking chains, and fully spelled-out identifier naming without domain-specific abbreviations), so that verbatim or line-by-line structural translation fails architectural gates by construction. Automated CI gates additionally verify standalone core builds, enforce one-way include boundaries that isolate optional add-ons from the core engine, and run unit-tested content and hashed-token leak checks on every pull request and public export to reject unapproved transcription phrasing, restricted source symbols or routine names from incompatible engines, and licensed third-party dataset payloads, ensuring that external reference datasets are resolved only at runtime while version control retains only format-level metadata and cryptographic hashes. Release packaging further inspects compiled symbol and link tables, bundles all required upstream license notices, emits a CycloneDX Software Bill of Materials (SBOM) with SHA-256 checksums and build-provenance attestations, and provides an optional BSD-3-Clause PocketFFT backend alongside FFTW3 so that serial binaries can be distributed free of copyleft runtime dependencies.
            \item \textbf{Regular Derivation Audits with AI Agents}: Because mechanical pattern and linter checks cannot by themselves detect subtler control-flow or structural carryover, Q3 supplements static gates with regular, version-pinned derivation audits conducted by specialized AI agents under a shared audit protocol. In the first pass, a comparison agent examines each Q3 module alongside the primary literature and any referenced public upstream routines at pinned commits, classifying every correspondence into a formal taxonomy: mathematical formulas, physical constants, interop-observable conventions, file-format facts, domain vocabulary, control-flow structure, and expression. Any structural similarity is permitted only when backed by an explicit \textit{dictated-by} justification demonstrating that the ordering or decomposition is forced by the mathematical recurrence, an external file-format specification, or a standard library contract (such as LAPACK, FFTW, or MPI semantics). In the second pass, an independent verifier agent performs a blind review followed by an adversarial challenge of both structural correspondences and kept-item classifications. Any routine whose structure is not strictly dictated by the underlying mathematics or interface is flagged for independent re-derivation from the published paper and re-verified against the black-box validation suite, while genuinely ambiguous cases are escalated to human maintainers with competing readings recorded neutrally rather than self-judged by an agent. All per-unit audit verdicts and consolidated ledgers are preserved as permanent evidentiary records.
            \item \textbf{Community Oversight}: Developing and publishing Q3 in the open, alongside documented provenance policies (\texttt{PROVENANCE.md} and \texttt{CONTRIBUTING.md}) and primary-literature citations in source headers, enables continuous inspection by the global electronic-structure community, including developers of established codes. Oversight is open to developers and maintainers of the foundational theories and established codes in the field and members of the broader electronic-structure ecosystems. Domain scientists may review physical formulations and inspect algorithmic decompositions and interface specifications (such as the native BerkeleyGW export and Wannier interpolation pipelines). External review of Q3 to date has been based on the published mathematical formalism, interface specifications (e.g., the BerkeleyGW export formats), and black-box numerical validation. At the organizational level, CAMSAI codifies provenance and scientific integrity as binding charter principles overseen by a vendor-neutral Steering Committee.
        \end{itemize}

    \subsection{Open Science, Consortium Governance, and Community Roadmap}
    \label{subsec:discussion-consortium}
    \label{subsec:discussion-future}

        To sustain Q3 and systematically address the remaining technical boundaries of the current release, the engine is developed and governed through the Consortium for the Advancement of Materials Science with AI (CAMSAI), an open community initiative welcoming participation from researchers across academia, industry, and the broader scientific community. Releasing the core foundation under the permissive Apache-2.0 license ensures that the codebase can be used, extended, and audited by the entire community without licensing friction, complementing open ecosystems such as AiiDA \cite{pizzi2016aiida}, JARVIS \cite{choudhary2020jarvis}, and Materials Project \cite{jain2013materialsproject}. Q3 is distributed as pre-built OCI container images (\texttt{ghcr.io/camsai/q3}), including serial and MPI variants for amd64 and arm64 and a CUDA variant for amd64, that run unmodified under Apptainer, podman-hpc, or Shifter for immediate deployment on HPC systems. For workflows that interface with proprietary simulation suites, commercially licensed pseudopotential datasets, or organization-specific formats, the build-time add-on architecture (\texttt{src/addons/<name>/}) allows licensed users and organizations to maintain and compile private format adapters locally without introducing proprietary dependencies, restricted dataset constants, or licensing constraints into the public repository.

        While Q3's core SCF engine, MPI pool scaling, and GPU backends are production-quality for semiconductors, metals, and collinear magnetic systems, several current architectural and physical limitations define the near-term development roadmap:
        \begin{itemize}
            \item \textbf{Intra-Pool Parallelism and Multi-GPU Scaling}: MPI k-point pool scaling is near-linear up to the irreducible k-point count on both CPUs and GPUs, but past that count additional ranks fall back to intra-pool parallelism. Currently, a $\Gamma$-only calculation cannot utilize multiple GPUs, and within a shared CPU pool the 3D FFT all-to-all redistribution spans the full grid rather than the cutoff sphere, allowing established plane-wave codes to remain faster at four or more ranks per pool (whereas Q3 leads at one and two ranks). Similarly, the 280-band benchmark system does not saturate an NVIDIA A100 ($4.1\times$ over a single CPU core, roughly matching 8 CPU threads). Ongoing development is addressing these bottlenecks by restricting FFT redistribution to the reciprocal-space cutoff sphere, adding NCCL collectives for multi-GPU band/G-vector decomposition within a single k-point pool, and validating AMD HIP and Intel SYCL backends on production GPUs.
            \item \textbf{Relativistic, Advanced Exchange-Correlation, and Excited-State Physics}: Extending Q3 beyond collinear LSDA/GGA+U requires two-component spinor wavefunctions for non-collinear magnetism and spin-orbit coupling (essential for topological and heavy-element materials), meta-GGA (e.g., SCAN \cite{sun2015scan} and r$^2$SCAN \cite{furness2020r2scan}) and screened hybrid functionals (HSE06 \cite{heyd2003hybrid, krukau2006hse}), and maximally localized Wannier functions \cite{marzari1997mlwf} with subspace disentanglement \cite{souza2001disentanglement} for entangled metallic bands. Because Q3's six-layer architecture isolates basis representations (Layer~3) and Hamiltonian operators (Layer~4) behind polymorphic interfaces, community contributors can develop and integrate these physical modules independently without modifying the underlying SCF or I/O infrastructure.
            \item \textbf{Broad Chemical Validation and Database Integration}: The current validation suite focuses on representative elemental and binary crystals (Si, Cu, Fe, Ni, Cr, SiC, diamond, FeO AFM+U, and graphite) and selected surface chemisorption and core-level spectroscopy benchmarks (CO and H on Ni(100)). Ongoing efforts are expanding automated regression benchmarks across transition-metal oxides, 2D heterostructures, and $f$-electron systems, while leveraging native CAMSAI standards \cite{camsai-standards, mat3ra-esse} schema compatibility to connect Q3 directly to high-throughput screening repositories \cite{choudhary2020jarvis, jain2013materialsproject}.
        \end{itemize}
        By aligning open-source governance with a modular, AI-legible codebase, CAMSAI provides a practical mechanism for the research community to convert these near-term limitations into validated capabilities.
    \section{Conclusion}
\label{sec:conclusion}

    This work has presented Quantum 3xplorer (Q3), a from-scratch quantum simulation engine demonstrating a new paradigm for \textbf{AI-driven quantum theory and computational materials science}, in which AI coding agents drive performant code generation, refactoring, and maintenance while allowing domain scientists to focus on high-level physical concepts, mathematical formulations, and rigorous numerical validation. Built around an \textit{AI-native} lifecycle (\textit{understandable for AI, written by AI, maintained by AI, and reviewed and validated by humans}), Q3 achieves full-stack plane-wave pseudopotential capabilities (norm-conserving, ultrasoft, and PAW), distributed MPI and multi-vendor GPU scaling on leadership HPC systems, and numerical parity with established reference implementations across semiconductors, metals, magnetic crystals, and adsorbate-covered surfaces in a matter of weeks of active development. We envision the pace of methodological and algorithmic development across the field to significantly accelerate as this human-supervised agentic workflow compresses the cycle required to translate new theoretical formulations from published equations into verified, production-grade parallel implementations. Combined with a method-agnostic core that cleanly separates quantum-mechanical operators from basis representations, schema-first data interoperability, native interfaces to many-body perturbation theory and Wannier interpolation, and multi-layered provenance assurance under the open-source CAMSAI consortium, Q3 provides a shared community foundation for unifying ground-state DFT, excited-state many-body methods, differentiable machine-learning workflows, and emerging quantum-classical algorithms.

    \section*{Acknowledgements}
    We used large language models (LLMs) to assist with drafting and editing portions of this manuscript. All content was reviewed and edited by the authors, who take full responsibility for the final text. J.V.\ acknowledges support by the U.S. Department of Energy, Office of Science, Office of Basic Energy Sciences, Chemical Sciences, Geosciences, and Biosciences Division, Catalysis Science Program to the Ultrafast Catalysis FWP 100435 at SLAC National Accelerator Laboratory under Contract DE-AC02-76SF00515. This research used resources of the National Energy Research Scientific Computing Center, a DOE Office of Science User Facility supported by the Office of Science of the U.S. Department of Energy under Contract DE-AC02-05CH11231.

    \section*{Author Contributions}
    T.B.\ conceived and supervised the project and led the development of the codebase and manuscript. V.B., S.V.B., and P.K.D.\ contributed to the data schemas, workflows, validation suites, and manuscript preparation. J.V.\ performed and analyzed the surface and X-ray photoelectron spectroscopy benchmarks. J.D.\ and M.D.B.\ advised on and performed benchmarking on NERSC Perlmutter and reviewed the BerkeleyGW interface and the manuscript; Lawrence Berkeley National Laboratory and the BerkeleyGW-affiliated co-authors are not developers or maintainers of Q3. F.H.J., Y.S., Z.L., S.G.L., K.C., R.P., and M.L.C.\ advised on the work and reviewed and edited the manuscript.

    \section*{Competing Interests}
    The authors declare no competing interests.
    
    \section*{Data and Software availability}
    The Quantum 3xplorer (Q3) source code is available as an open-source project under the Apache-2.0 license at \href{https://github.com/camsai/q3}{github.com/camsai/q3}. Pre-built container images (serial and MPI for amd64 and arm64, CUDA for amd64) are published at \texttt{ghcr.io/camsai/q3}. The CAMSAI standards and associated schemas are maintained at \href{https://github.com/camsai/standards}{github.com/camsai/standards} \cite{camsai-standards, mat3ra-esse}.

    \bibliographystyle{unsrtnat}
    \bibliography{references}

\end{document}